\documentclass[twocolumn,prl]{revtex4}
\usepackage{graphicx}
\usepackage{amsfonts}
\usepackage{txfonts}

\newcommand{\beq}{\begin{equation}}
\newcommand{\eeq}{\end{equation}}
\newcommand{\beqa}{\begin{eqnarray}}
\newcommand{\eeqa}{\end{eqnarray}}
\newcommand{\beqar}{\begin{eqnarray*}}
\newcommand{\eeqar}{\end{eqnarray*}}

\def\bi{\begin{itemize}}
\def\ei{\end{itemize}}
\def\be{\begin{equation}}
\def\ee{\end{equation}}
\def\bea{\begin{eqnarray}}
\def\eea{\end{eqnarray}}
\def\ben{\begin{eqnarray*}}
\def\een{\end{eqnarray*}}

\def\>{\rangle}
\def\<{\langle}

\newcommand{\mE}{{\mathcal E}}
\newcommand{\mU}{{\mathcal U}}
\newcommand{\mX}{{\mathcal X}}
\newcommand{\mY}{{\mathcal Y}}
\newcommand{\mZ}{{\mathcal Z}}
\newcommand{\mG}{{\mathcal G}}

\newcommand{\mH}{{\mathcal H}}
\newcommand{\mI}{{\mathcal I}}

\newcommand{\1} I %{{\openone}}

\newcommand{\bra}[1]{\langle #1 |}
\newcommand{\ket}[1]{| #1 \rangle}

\def\*{\star}

\def\tilde{\widetilde}

\begin{document}

\title{Graphical Quantum Error-Correcting Codes}

\author{Sixia Yu$^{1,2\footnote{Email: yusixia@ustc.edu.cn}}$, Qing Chen$^1$, and Choo Hiap Oh$^2$}
\affiliation{$^1$Hefei National Laboratory for Physical Sciences at
Microscale and Department of Modern Physics, University of Science
and Technology of China, Hefei 230026, P.R. China \\
$^2$Department of Physics, National University of Singapore, 2
Science Drive 3,  Singapore 117542 }
\date{\today}


\begin{abstract}
We introduce a purely graph-theoretical object, namely \textit{the
coding clique}, to construct  quantum error-correcting codes. Almost
all quantum codes constructed so far are stabilizer (additive) codes
and the construction of nonadditive codes, which are potentially
more efficient, is not as well understood as that of stabilizer
codes. Our graphical approach provides a unified and classical way
to construct both stabilizer and nonadditive codes. In particular we
have explicitly constructed the optimal ((10,24,3)) code and a
family of 1-error detecting nonadditive codes with the highest
encoding rate so far. In the case of stabilizer codes a thorough
search becomes tangible and we have classified all the extremal
stabilizer codes up to 8 qubits.
\end{abstract}
\maketitle

\section*{Introduction} Quantum error correcting code (QECC)
\cite{shor,bdsw,kl,steane} has become an indispensable element in
many quantum informational tasks such as the fault-tolerant quantum
computation \cite{ftc}, the quantum key distributions \cite{qkd},
and the entanglement purification \cite{ep} and so on, to fight the
noises. Roughly speaking, a QECC is a subspace of the Hilbert space
of a system of many qubits (physical qubits) with the property that
the quantum data (logical qubits) encoded in this subspace can be
recovered faithfully, even though some physical qubits may suffer
arbitrary errors, by a measurement followed by suitable unitary
transformations.

Almost all QECCs constructed so far are stabilizer or additive codes
\cite{gott1,crss,calder} whose code space is specified by the joint
+1 eigenspace of a stabilizer, an Abelian group of tensor products
of Pauli operators. Special examples include the CSS codes
\cite{css}, the topological codes \cite{tc1}, color codes
\cite{tc2}, and the recently introduced entanglement-assisted codes
\cite{eaqecc}. Because of its structure a stabilizer code has always
a dimension that is a power of 2 and is usually denoted as
$[[n,k,d]]$, which is a $2^k$ dimensional subspace of an $n$-qubit
Hilbert space. Here $d$ denotes the distance of a quantum code
meaning  that $[\frac{d-1}2]$-qubit errors can be corrected and all
the errors that acts nontrivially on less than $d$ qubits either can
be detected or stabilize the code subspace. Apart from a
classification of the $[[n,0,d]]$ codes up to 12 qubits
\cite{sdual}, the classification of stabilizer codes encoding a
nonzero number of logical qubits has not been achieved yet.

The code without a stabilizer structure is called as a {\it
nonadditive code}. Usually we denote by $((n,K,d))$ a nonadditive
QECC of distance $d$ that is a $K$-dimensional subspace of an
$n$-qubit Hilbert space correcting $[\frac{d-1}2]$-qubit. Since less
structured than the additive codes, the nonadditive codes may be
more efficient in the sense of  a larger code subspace on one hand,
harder to be constructed on the other hand.  The first nonadditive
code \cite{rains1} that outperforms the stabilizer codes is the
1-error detecting code $((5,6,2))$. Its generalization \cite{rains2}
is outperformed by another family of codes of distance 2
\cite{smolin}. There are also some efforts to construct nonadditive
codes systematically \cite{nonadd,nonadd2,calder2}. Only recently
the first error-correcting code, namely a ((9,12,3)) code, that
outperforms the optimal stabilizer code is explicitly constructed in
\cite{12in9}. Later on some other nonadditive codes have also been
found \cite{zb}.

In this article we present a graphical approach to construct  both
additive and nonadditive codes in a unified and systematic manner.
Graph is related intimately to the construction of classical
error-correcting codes (e.g. Tanner graph \cite{tanner}) and also
finds applications in the quantum stabilizer code via graph states
\cite{graph,werner,schling,grassl}. With the help of graph states,
we relate a pure graph-theoretic object
--- the coding cliques --- to the construction of quantum codes,
especially of nonadditive codes. An algorithm to search
systematically for quantum codes via coding cliques is outlined.

\section*{Graph and graph-state basis} A\textit{ graph}
$G=(V,\Gamma)$ is composed of a set $V$ of $n$ vertices and a set of
edges specified by the {\it adjacency matrix} $\Gamma$, which is an
$n\times n$ symmetric matrix with vanishing diagonal entries and
$\Gamma_{ab}=1$ if vertices $a,b$ are connected and $\Gamma_{ab}=0$
otherwise. Here we consider only undirected simple graphs. The {\it
neighborhood of a vertex} $a$ is denoted by $N_a=\{v\in
V|\Gamma_{av}=1\}$, i.e, the set of all the vertices that are
connected to $a$.

The graph states \cite{werner,graph} are useful multipartite
entangled states that are essential resources for the one-way
computing \cite{oneway} and can be experimentally demonstrated
\cite{six}. Also the graph state plays the key role in our graphical
construction of QECC. Consider a system of $n$ qubits that are
labeled by those $n$ vertices in $V$ and denote by $\mathcal
X_a,\mathcal Y_a, \mathcal Z_a$, and $\mI_a$ three Pauli operators
and identity matrix acting on the qubit $a\in V$. The $n$-qubit {\it
graph state} associated with $G$ reads \cite{werner,graph}
\begin{equation}
\ket \Gamma=\prod_{\Gamma_{ab}=1}\mU_{ab}|+\rangle
^V_x=\frac1{\sqrt{2^n}}\sum_{\vec\mu=\bf 0}^{\bf 1}
 (-1)^{\frac12\vec\mu\cdot\Gamma\cdot\vec\mu}\ket{\vec\mu}_z,
\end{equation}
where $\ket{\vec\mu}_z$ is the joint eigenstate of $\mathcal Z_a$
$(a\in V)$ with $(-1)^{\mu_a}$ as eigenvalues, $|+\rangle_x^V$ is
the joint +1 eigenstate of $\mX_a$ ($a\in V$), and $\mU_{ab}$ is the
controlled phase gate between qubits $a$ and $b$. The {\it
graph-state basis} of the $n$-qubit Hilbert space $\mH_n$ refers to
$\{\ket {\Gamma_C}:=\mathcal Z_{C}\ket \Gamma\}$ with $C\in 2^V$,
the set of all the subsets of $V$. A collection of subsets
$\{C_1,C_2,\ldots,C_K\}$ specifies a $K$ dimensional subspace of
$\mH_n$ that is spanned by the graph-state basis
$\{\ket{\Gamma_{C_i}}\}_{i=1}^K$. The graph state $\ket\Gamma$ is
also the unique joint +1 eigenstate of $n$ {\it vertex stabilizers}
\begin{equation}\label{vs}
\mathcal G_a=\mathcal X_a\prod_{b\in N_a}\mathcal Z_b:=\mathcal
X_a\mathcal Z_{N_a}, \quad a\in V.
\end{equation}
Obviously $\mG_a|\Gamma_C\rangle=-|\Gamma_C\rangle$ if $a\in C$ and
$\mG_a|\Gamma_C\rangle=|\Gamma_C\rangle$ otherwise. For a given
subset $S\subseteq V$ the operator $\mathcal G_S:=\prod_{a\in
S}\mathcal G_{a}$ also stabilizes the graph state, i.e.,
$\mG_S\ket\Gamma=\ket\Gamma$.

A graph determines uniquely a graph state and two graph states
determined by two graphs are equivalent up to some \textit{local
Clifford transformations} (LCTs) iff these two graphs are related to
each other by \textit{local complements} (LCs) \cite{LC}. An LCT is
a special local unitary transformation that maps Pauli operators to
themselves up to some sign changes. An LC of a graph on a vertex $v$
refers to the operation that in the neighborhood of $v$ we connect
all the disconnected vertices and disconnect all the connected
vertices. All the graphs on up to 12 vertices have been classified
under LCs and graph isomorphisms \cite{sdual}.

\section*{Coding cliques and QECC} From the standard theory of the
QECC we know that if  a set of Pauli errors can be corrected then
all the errors that are linear combinations of the Pauli errors
belonging to that set can also be corrected. One crucial observation
is that when acting on the graph states, because of the vertex
stabilizers in Eq.(\ref{vs}), the Pauli errors can be equivalently,
up to some signs, replaced by phase flip errors, which motivates us
to introduce the notion of coding cliques as follows.

Given a graph $G=(V,\Gamma)$ we define the {\it neighborhood of a
subset} $S\subseteq V$ to be $N_S=\bigtriangleup_{v\in S}N_v$, where
we have denoted by $C\bigtriangleup D:=C\cup D-C\cap D$ the
\textit{symmetric difference} of two sets $C,D\subseteq V$. Given
any integer $d$ ($1\le d\le n)$ we define a {\it $d$-purity set} as
\begin{equation}
\mathbb{S}_d=\left\{S\subseteq V\Big|\;|S \cup N_S|< d\right\}.
\end{equation}
Here $|S|$ denotes the number of the elements in set $S$. For a
subset of vertices $C\subseteq V$ if there exist two subsets
$\omega,\delta\subseteq V$ such that $C=\delta\bigtriangleup
N_\omega$ and $|\omega\cup\delta|=p$, we call the ordered pair
$(\delta,\omega)$ as a {\it $p$-cover} of $C$ or equivalently $C$ is
said to be {\it covered} by $p$ vertices or by $(\delta,\omega)$. We
denote by $\mathbb D_d$ the \textit{$d$-uncoverable set} that
contains all the subsets of $V$ which cannot be covered by less than
$d$ vertices, i.e.,
\begin{equation}
\mathbb{D}_d=2^V-\left\{\delta\bigtriangleup
N_\omega\Big|\;|\omega\cup\delta|< d\right\}.
\end{equation}
A \textit{coding clique} $\mathbb C_d^K$ of a graph $G$ is a
collection of vertex subsets $\{C_1,C_2,\ldots,C_K\}$ that satisfies
\begin{description}
\item[Condition 0.] $\emptyset\in \mathbb C_d^K$;
\item[Condition 1.] $|C_i\cap S|$ is even for all $1\le i\le K$ and $S\in \mathbb
S_d$;
\item[Condition 2.] $C_i\bigtriangleup C_j\in \mathbb
D_d$ for all $1\le i\ne j\le K$.
\end{description}

A $K$-clique of a given graph refers to a subset of $K$ vertices
that are pairwise connected. Our coding clique $\mathbb C_d^K$ of a
graph $G$ is exactly a $K$-clique of the\textit{ super graph}
$\mathbb G$ defined as follows. The vertices of the super graph
$\mathbb G$ include the empty set and all the nonempty vertex
subsets that belong to $\mathbb D_d$ and satisfy Condition 1. Two
different vertices $C,C^\prime$ of the super graph $\mathbb G$ are
connected if $C\bigtriangleup C^\prime \in \mathbb{D}_d$. Thus the
coding clique $\mathbb C_d^K$ is a pure graph-theoretic object that
is formulated in a constructive way. The following theorem relates
the coding cliques to the construction of QECCs. (See Appendix for
proofs.)

{\bf Theorem 1} {\it Given a coding clique $\mathbb C_d^K$ of a
graph $G$ on $n$ vertices, the subspace spanned by the graph-state
basis $\{\ket{\Gamma_C}|C \in \mathbb{C}_{d}^K\}$, denoted as
$(G,K,d)$, is an $((n,K,d)) $ code.}

As an application, a systematic search for the quantum codes can be
done according to the following algorithm: i) To input a graph
$G=(V,\Gamma)$ on $n$ vertices that may be connected or
disconnected; ii) To choose a distance $d$ and compute the
$d$-purity set $\mathbb S_d$ and the $d$-uncoverable set $\mathbb
D_d$ so that a super graph $\mathbb G$ can be built; iii) To find
all the $K$-clique $\mathbb C_d^K$ of the super graph $\mathbb G$;
iv) For every coding clique we obtain a $(G,K,d)$ code , i.e., an
$((n,K,d))$ code that is spanned by the graph-state basis
$\{\ket{\Gamma_C}=\mZ_C\ket\Gamma|C\in\mathbb C_d^K\}$.

It is not difficult to see that if the $d$-purity set $\mathbb S_d$
is empty then the $(G,K,d)$ code is pure, i.e., every error acting
nontrivially on less than $d$ qubits can be detected.

If a coding clique form a group with respect to the symmetric
difference $\bigtriangleup$, then this coding clique is referred to
as a \textit{coding group} of the graph. Because the coding group is
an Abelian group with self inverse, the number $K$ of its elements
must be a power of 2, i.e., $K=2^k$ for some integer $k$, which is
referred to as the dimension of the coding group. A $k$-dimensional
coding group has $k$ independent generators. If we find a coding
group then we obtain a stabilizer code and all the stabilizer codes
can be found this way because of the following theorem.

{\bf Theorem 2} {\it Every coding group  $\mathbb C_d^{2^k}(G)$
provides a stabilizer code $[[n,k,d]]$,  denoted as a $[G,k,d]$
code. Every stabilizer code $[[n,k,d]]$ is equivalent to a $[G,k,d]$
code for some graph $G$ on $n$ vertices.}

From the same graph we may obtain inequivalent $(G,K,d)$ codes and
different graphs may provide equivalent codes. To reduce the number
of the graphs to be searched we have also investigated how the
coding clique changes under the local complements of the graph. For
convenience we denote by $G_v$ the graph obtained from $G$ by making
an LC on vertex $v$. For a given vertex $v$ and a subset $C\subseteq
V$ we denote $C_v=C$ if $v\notin C$ and $C_v=C\bigtriangleup N_v$ if
$v\in C$.

\textbf{LC rule for coding cliques} \textit{If $\mathbb C^K_d$ is a
coding clique (group) of the graph $G$, then $\tilde{\mathbb
C}^K_d=\{C_v|C\in\mathbb C^K_d\}$ is the coding clique (group) of
the graph $G_v$. Two codes specified by coding cliques (groups)
$\mathbb C^K_d$ and $\tilde{\mathbb C}^K_d$ of $G$ and $G_v$ are
equivalent under LCTs.}

Two quantum codes  are regarded to be equivalent if they are related
to each other by LCTs plus permutations of physical qubits.
Therefore we need only to take those inequivalent classes of graphs
under LCs and graph isomorphisms as inputs to our algorithm. In what
follows we shall document some results obtained via the algorithm
outlined above \cite{cliquer}.  A QECC will be specified by a graph
together with a coding clique. For a stabilizer code we will only
specify the generators of the coding group and for a nonadditive
code we will list all the members of the coding clique.

\section*{Nonadditive codes} At first let us reproduce some known
nonadditive codes via our graphical construction. The first example
is the first nonadditive 1-error detecting code $((5,6,2)$)
\cite{rains1} that outperforms the optimal stabilizer code
$[[5,2,2]]$. The graph provides the code is the loop graph $L_5$ on
5 vertices as shown in Fig.1.(a) and the coding clique is
$\{C_i\}_{i=1}^6$ where
\begin{eqnarray} \label{ci}
&C_1=\emptyset,\;C_2=\{235\},\,C_3=\{341\},&\cr &C_4=\{452\},\,
C_5=\{513\},\;C_6=\{124\}.&
\end{eqnarray}
The second example comes from the first 1-error correcting
nonadditive code $((9,12,3))$ that outperforms the optimal
stabilizer code $[[9,3,3]]$. The code is specified by the loop graph
$L_9$ on 9 vertices and the coding clique contains those 12 subsets
$\{V_i\}_{i=1}^{12}$ defined as in \cite{12in9}. Our search results
show that the code $(L_5,6,2)$  and the code $(L_9,12,3)$ are
unique. In addition there is no $(G,13,3)$ code for any graph $G$ on
9 vertices.
\begin{figure}
\begin{center}\includegraphics[scale=0.7]{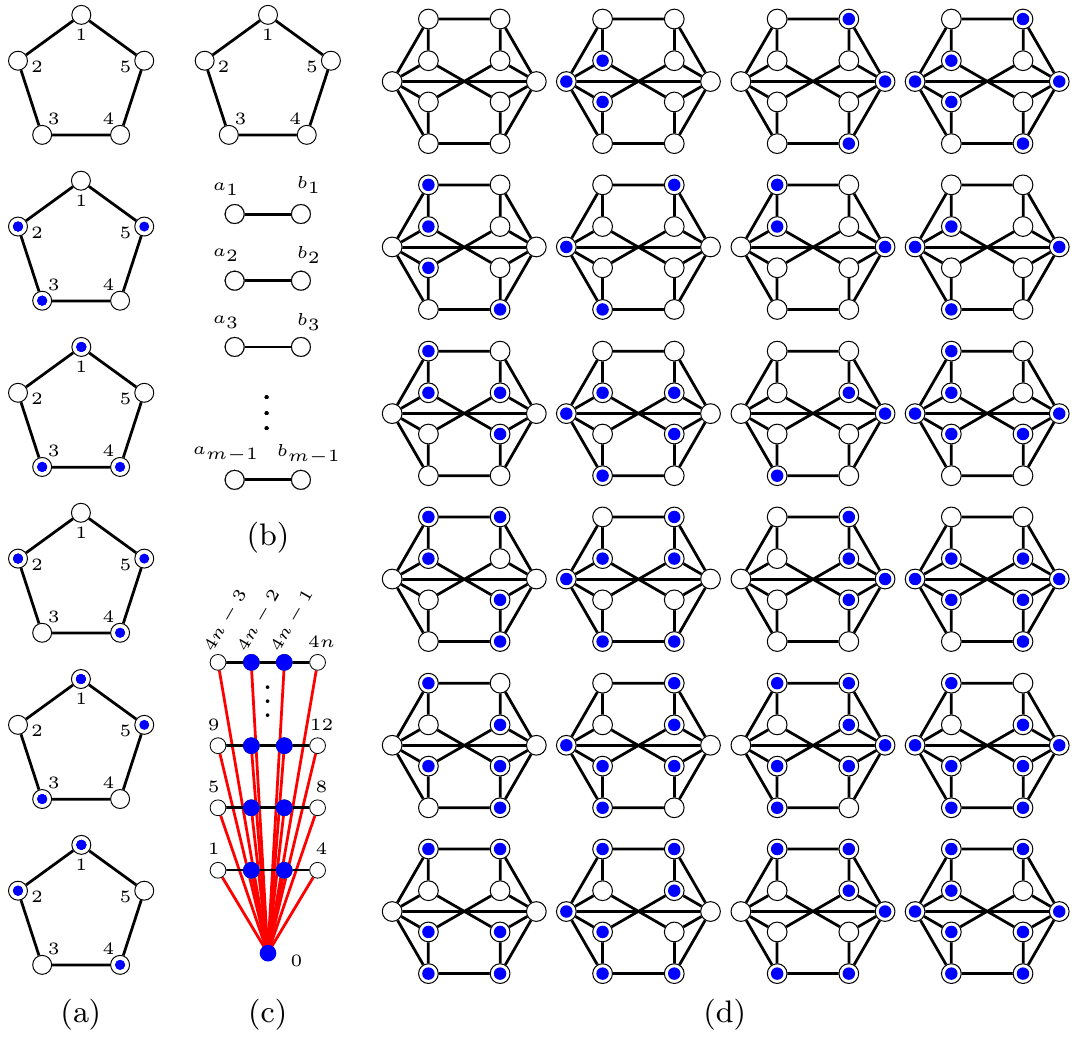}
\caption{The graphs and coding cliques for nonadditive codes. (a)
Six elements of the coding clique for the $((5,6,2))$ code. (b) A
graph for Rain's $((2m+3,6\cdot 4^{m-1},2))$ code. (c) The graph for
the $((4n+1,M_n+1,2))$ code together with an element of the coding
clique. (d) The $((10,24,3))$ code. }\end{center}
\end{figure}

Another example is Rain's $((2m+3,6\cdot 4^{m-1},2))$ code
\cite{rains2}, which can be specified by the graph made up of $L_5$
and $2(m-1)$ pairwise connected vertices as shown in Fig.1(b). The
coding clique contains all the subsets of form $\{C_i\bigtriangleup
U\}_{i=1}^6 $ where $C_i$'s are given in Eq.(\ref{ci}) and $U$ can
be any one of $2^{2m-2}$ subsets of vertices generated by
$\{2,a_j\}$ and $\{5,b_l\}$ for $j,l=1,2,\ldots,m-1$ with respect to
the symmetric difference $\bigtriangleup$.

Now let us construct some new codes. The first result is a family of
1-error detecting codes $((4n+1,M_n+1,2))$ where
$M_n=2^{4n-1}-\frac12C_{4n}^{2n}$. A family of nonadditive codes of
distance 2 \cite{smolin}  is constructed which encodes an
$M_n$-dimensional subspace if there are $4n+1$ physical qubits. The
code can be specified by the star graph on $4n+1$ vertices
$V_{4n+1}$ centered on vertex $o$ as shown in Fig.1 (c) (indicated
by red edges). The coding clique reads
\begin{equation}\label{ci2}
\left\{C\subset V_{4n+1} \Big| o\notin C;|C|=2l,\; \mbox{or}
\;|C|=2n+2l+1, 0\le l\le n-1\right\}.
\end{equation} For the star
graph the dimension $M_n$ of the code space is optimal
\cite{smolin}. We consider instead the graph as shown in Fig.1(c)
that is built on a star graph with red edges and some additional
black edges. The coding clique, in addition to Eq.(\ref{ci2}),
contains one more subset of all $2n+1$ blue-colored vertices
$\{o,4l-2,4l-1\}_{l=1}^n$ as indicated in Fig.1(c). Therefore we
obtain a family of nonadditive code of distance 2 with an encoding
rate $M_n+1$.

The second result is the optimal $((10,24,3))$ code. The linear
programming bound \cite{calder} indicates  that the largest code
subspace of a 1-error correcting code on 10 qubits  is of dimension
24. Among 3132 inequivalent connected graphs and all the
disconnected graphs on 10 vertices there is a unique $(G_{10},24,3)$
code as depicted in Fig.1(d) in which blue-colored vertex sets
indicate all the nonempty elements of the coding clique.

\section*{Stabilizer codes} Because of Theorem 2 the search on all
the inequivalent graphs on $n$ vertices for all the coding groups
according to our algorithm provided above will exhaust all the
stabilizer codes of $n$ qubits. Therefore a classification of all
stabilizer codes is tangible. We consider in the following the
classification of those extremal codes, i.e., the code with a
maximal $k$ or a maximal $d$ given $n$, up to 8 qubits, as listed in
the table in \cite{calder}.

\begin{figure}
\begin{center}\includegraphics[scale=0.8]{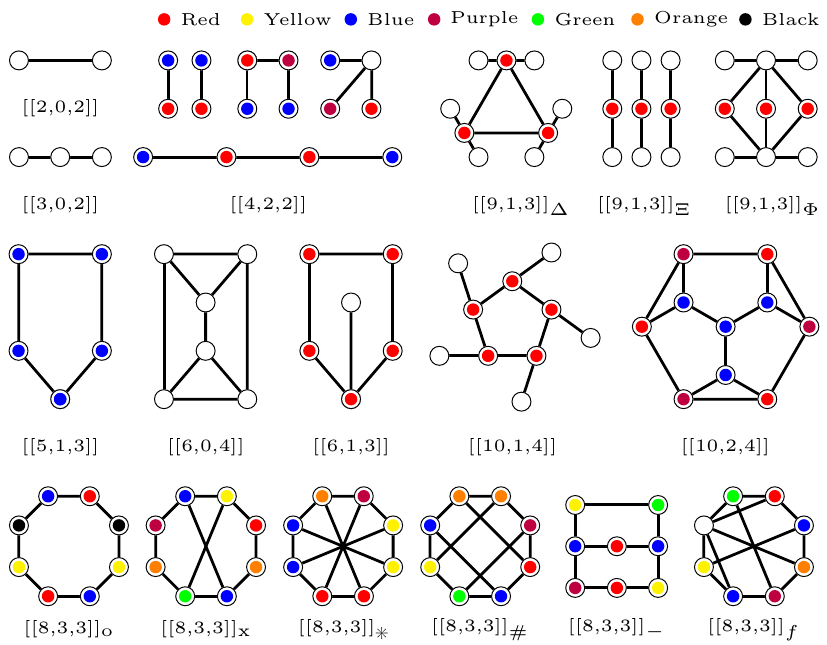}
\caption{The classification of stabilizer codes part I. There are
three primary colored sets of vertices: red set $R$, yellow set $Y$,
and blue set $B$. Orange, green, purple, and black vertices belong
to $R\cap Y$, $Y\cap B$, $B\cap R$, and $R\cap Y\cap B$,
respectively. Each primary color set represents a generator of the
coding group.}\end{center}
\end{figure}

Up to six qubits, there are three extremal codes $[[4,2,2]]$,
$[[5,1,3]]$, and $[[6,1,3]]$, which are all unique and corresponding
graphs and coding groups are shown in Fig.2. For $[[4,2,2]]$ we have
shown 4 different graphs and coding groups, all resulting an
equivalent code. There is also a trivial $[[6,1,3]]$ code obtained
from $[[5,1,3]]$ code by appending an isolated vertex to the
pentagon as noticed in \cite{calder}. In the classification of
7-qubit codes we will neglect also those codes obtained from
$[[5,1,3]]$ and $[[6,1,3]]$ by appending a disconnected subgraph.
\begin{figure}
\begin{center}\includegraphics[scale=0.8]{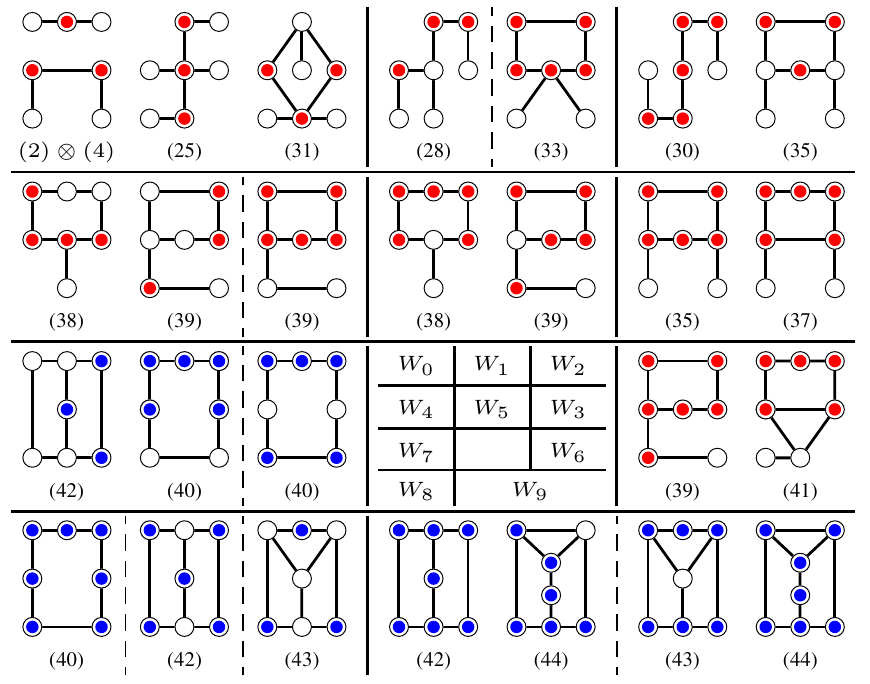} \caption{
Classification of stabilizer codes part II. The table of 16
inequivalent $[[7,1,3]]$ codes with 10 different weight
distributions separated by lines and dashed lines. Graphs with blue
vertices provide pure codes and graphs with red vertices provide
impure codes. }
\end{center}
\end{figure}

Among 26 inequivalent connected graphs and on 7 vertices there are
14 graphs that admit a 1-dimensional coding group as shown in Fig.3,
where the numberings of the inequivalent graphs are adopted from
\cite{graph}. One graph may provide inequivalent codes and different
graphs may provide equivalent codes. Steane's 7-qubit code
\cite{steane} is provided by graph No.43 with the weight
distribution $W_8$. (A detailed account of weight distribution
\cite{enum} is given in Appendix.) It is interesting to notice that
a $[[7,1,3]]$ code can be provided by a disconnected graph, whose
two subgraphs cannot admit any coding clique alone. Disconnected
graphs on 7 vertices provides no further $[[7,1,3]]$ codes.

Among 101 inequivalent connected graphs on 8 vertices there are only
6 LU-inequivalent graphs that admit 3-dimensional coding groups as
shown in Fig.2. In each graph there may be many different coding
groups and only one coding group is shown for each graph.
Disconnected graphs on 8 vertices provides no further $[[8,3,3]]$
codes. From the weight distributions of corresponding codes it can
be proved that the code $[[8,3,3]]$ is unique (Appendix).  In Fig.2
we also show three different graphs and the coding groups that
provide codes that are equivalent to Shor's $[[9,1,3]]$ code
\cite{shor} and a LC-equivalent graph for the impure $[[10,1,4]]$
code constructed in \cite{werner}. In comparison we have found a
pure $[[10,2,4]]$ code with a weight distribution
$(90_6,135_8,30_{10})$.

\section*{Discussion} We provide a classical recipe to cook quantum
codes, which is exhaustive for stabilizer codes and systematic for
nonadditive codes. A classification of stabilizer codes has been
done up to 8 qubits. Since all the known good nonadditive codes can
be reproduced and an optimal code ((10,24,3)) is constructed via our
graphical approach, we speculate that our algorithm is also
exhaustive for nonadditive codes.

With an improved computational power a classification of the
stabilizer codes on more qubits and the discovery of more good codes
either stabilizer or nonadditive codes are expected. It should
admitted that the clique finding problem is intrinsically a
NP-complete problem. Therefore analytical constructions of the
coding cliques, at least for some special family of graphs such as
loop graphs and hyper cubic graphs, deserve exploring. A direct
construction of a series nonadditive 1-error detecting codes with
highest encoding rate so far described here may provide a clue. In
addition it is not difficult to generalize the idea of coding
cliques to the construction of nonbinary codes via graph states for
systems with more than 2 levels.

Our precious quantum data can be protected from decoherences either
in a dynamical manner as in the decoherence-free subspace approach,
or in a geometric manner as in geometric computations, or in a
topological manner as in the topological quantum computations based
on the topological codes, a special stabilizer codes. Our results
make a bridge between the exciting classical field of graph theory
and the quantum error correction. Equipped with the one-way
computation model based on graph states and the graphical QECCs, we
may envision a graphical quantum computation based directly on the
graphical objects.

SXY acknowledges financial support from NNSF of China (Grant No.
90303023 and Grant No. 10675107), CAS, and WBS (Project Account No):
R-144-000-189-305, Quantum information and Storage (QIS).

\section*{Appendix}
\paragraph*{Proof of Theorem 1}
It is enough to prove that for any error $\mE_d$ acting nontrivially
on less than $d$ qubits we have
$\bra{\Gamma_C}\mE_d\ket{\Gamma_{C^\prime}}=f(\mE_d)\delta_{CC^\prime}$
\cite{kl,werner} for all $C,C^\prime\in \mathbb C_d^K$. Without lose
of generosity we assume that $\mE_d=\mathcal X_\omega \mathcal
Z_\delta$ for some pair of subsets $\delta,\omega$ with
$|\omega\cup\delta|< d$, which represents that there are $\mX$,
$\mY$, and $\mZ$ errors on the qubits in $\omega-\delta\cap\omega$,
$\omega\cap \delta$, and $\delta-\omega\cap\delta$ respectively.
When acting on the graph state the error $\mE_d\propto\mathcal
G_\omega \mathcal Z_\Omega$ can be replaced by phase flip errors
$\mZ_\Omega$ on $\Omega:=\delta\bigtriangleup N_\omega$. If $\Omega$
is empty then $\delta=N_\omega$ and the error is proportional to
$\mathcal G_\omega$. In this case we have
$\mG_\omega|\Gamma_C\rangle=|\Gamma_C\rangle$ for all $C\in \mathbb
C_d^k$ because $|\omega\cap C|$ is even which stems from the fact
that $|\omega\cup N_\omega|<d$, i.e., $\omega\in \mathbb{S}_p$ and
Condition 1. Thus the error behaves like a constant operator on the
coding subspace and can be neglected. If $\Omega$ is not empty then
$\Omega\notin\mathbb D_d\cup \emptyset$ because it is covered by
$(\delta,\omega)$ and $|\delta\cup\omega|<d$. As a result
$\bra{\Gamma_C}\mE_d\ket{\Gamma_{C^\prime}}\propto\bra{\Gamma}
\mZ_{C\bigtriangleup C^\prime}\mZ_\Omega\ket\Gamma=0$ for all
$C,C^\prime\in \mathbb C_d^k\cup \emptyset$ because condition 2
ensures that $C\bigtriangleup C^\prime\ne \Omega$.

\paragraph*{Proof of Theorem 2} We prove the second part first, i.e. to
construct a graph and a coding clique from a given stabilizer code.
We do not need to introduce those input vertices as in
Ref.\cite{werner}.

A stabilizer is a set of commuting observables that are tensor
products of Pauli operators $\{\mathcal X,\mathcal Y,\mathcal Z\}$
and the identity operator $\mathcal I$ on each qubit. A stabilizer
code $[[n,k,d]]$ is the simultaneous +1 eigenspace of $n-k$
generators $\langle\mathcal S_1,\mathcal S_2, \ldots, \mathcal
S_{n-k}\rangle$ of the stabilizer. Stabilizer codes that can be
related to each other by a local Clifford transformation (LCTs) and
permutations of physical qubits are equivalent.

With the help of the binary representation  $(\mathcal I\to 00$,
$\mathcal X\to 10$, $\mathcal Z\to 01$, $\mathcal Y\to 11)$ of the
local operators we can describe the stabilizer a check matrix
$[G_x|G_z]$ where $(n-k)\times n$ matrix $G_{x(z)}$ is formed by all
the first (second) digits of the local operators of the generators
of the stabilizer. By relabling or a different choice of the
generators of the stabilizer, permutating the qubits, and making
suitable LCTs, one can always bring the check matrix into its
standard form \cite{n-r}
\begin{equation}\label{sf}
[I_{r},A_{r\times k}|D_{r\times r}+AE^T,E_{r\times k}]\quad (r=n-k)
\end{equation}
where $I_{r}$ denotes the identity matrix and the submatrix $D$ is
symmetric as required by the commutativity among the generators,
i.e., $G_xG_z^T$ being symmetric. Here all the additions are
additions module 2.

A basis of the code subspace can be constructed by adding to the
stabilizer additional $k$ independent generators. The most general
choice, up to an LCT and a choice of different generators, is to add
the $k$ generators specified by the check matrix
\begin{equation}\label{gc}
[0,I_k|E^T+FA^T,F_{k\times k}]
\end{equation}
where $F$ is an arbitrary symmetric matrix. By adding the check
matrix Eq.(\ref{gc}) multiplied by $A$ from left to the the standard
form Eq.(\ref{sf}) we obtain the check matrix of $n$ generators as
\begin{equation}
\left[\matrix{I_r&0\cr 0&I_{k}}\right| \left.\matrix{D+AFA^T&E+AF\cr
E^T+FA^T&F}\right]:=[I_n|\Gamma].
\end{equation}
If one of the diagonal element $\Gamma$, say $\Gamma_{mm}$, is
nonzero then we can make it vanish by performing a LCT $(\mathcal
Y\to \mathcal X\to -\mathcal Y; \mathcal Z\to \mathcal Z)$ to the
$m$-th qubit. Therefore we obtain a standard adjacency matrix
$\Gamma$ of a graph: $n\times n$ symmetric matrix with vanishing
diagonal entries. In terms of the vertex stabilizers $\{\mathcal
G_a\}$ of this graph, the stabilizer is generated by
\begin{equation}
\mathcal G_a\prod_{c=1}^k (\mathcal G_{n-k+c})^{A_{ac}} \quad
a=1,\ldots,n-k.
\end{equation}

Denote by $C_m$ the set of vertices on which the entries of the
$m$-th row $(m=1,\ldots,k)$ of the $k\times n$ matrix $[A^T,I_k]$ do
not vanish and by $\mathbb C$ the set of subset of $V$ generated by
$\{C_m\}_{m=1}^k$ with respect to the symmetric difference
$\bigtriangleup$. Then the graph-state basis $\{\mathcal Z_{C}\ket
\Gamma|C\in\mathbb C \}$ spans the $[[n,k,d]]$ code. By definition
$\mathbb C$ is a group with respect to the symmetric difference and
$\emptyset\in \mathbb C$. Secondly for every $S\in\mathbb{S}_d$ the
graph stabilizer $\mathcal G_S$, which can be a possible error, must
stabilize the code subspace, i.e., $|S\cap C|$ must be even for all
$C\in \mathbb C$. Thirdly, had one nonempty $C\in \mathbb C$ a
$p$-cover $(\omega,\delta)$ with $p<d$, the error operator $\mathcal
X_{\omega}\mathcal Z_{\delta}$ would be incorrigible because
$\mathcal X_{\omega}\mathcal Z_{\delta}\ket G\propto\mathcal Z_C\ket
G$. Thus $\mathbb C\subseteq \mathbb D_d$, i.e., Condition 3 is
satisfied. As a result $\mathbb C$ is a coding group.

One the other hand if we have a $k$ dimensional coding group of a
graph $G$ which is generated by $\langle C_1,C_2,\ldots,C_k\rangle$,
then we have a code $(G,2^k,d)$ according to Theorem 1. Since $k$
constraints $|S\cup C_i|=$even for $i=1,2,\ldots,k$ have exactly
$n-k$ independent solutions which are denoted as $\langle
S_1,S_2,\ldots,S_{n-k}\rangle$, we obtain a stabilizer of the code
that is generated by $\langle\mG_{S_i}\rangle_{i=1}^{n-k}$.

\paragraph{Proof of LC rule for coding cliques}
Given a graph $G=(V,\Gamma)$ and a coding clique $\mathbb C_d^K$ we
denote $\tilde{\mathbb C}_d^K=\{C_v|C\in \mathbb C_d^K\}$ with
$C_v=C$ if $v\notin C$ and $C_v=C\bigtriangleup N_v$ otherwise where
$N_v$ is the neighborhood of $v$ in the graph $G$. For convenience
we denote $A\bigtriangleup^mB=A$ if $m$ is even and
$A\bigtriangleup^mB=A\bigtriangleup B$ if $m$ is odd. Then we have
concisely  $C_v=C\bigtriangleup^{|v\cap C|}N_v$, where we have
denoted the single vertex set containing $v$ also by $v$. After a
local complement on vertex $v$ made to $G$ we obtain a new graph
$G_v=(V,\tilde \Gamma)$. The neighborhood of a vertex set
$S\subseteq V$ in the new graph $G_v$ reads
\begin{equation}
\tilde N_S=N_S\bigtriangleup (S\cap N_v)\bigtriangleup^{|S\cap
N_v|}N_v
\end{equation}
where $N_S$ is the neighborhood of $S$ in the original graph $G$.

Firstly since $\emptyset\in \mathbb C_d^K$ and $v\notin \emptyset$
we have $\emptyset\in \tilde{\mathbb C}_d^K$. Condition 0 is
satisfied. Secondly, by denoting  $S_v=S\bigtriangleup^{|S\cap
N_v|}v$ we see that if $S\in\mathbb S_d(G_{v})$, i.e., $S$ belongs
to the $d$-purity set of $G_v$ then  $S$ belongs to the $d$-purity
set of $G$,  i.e., $S_v\in \mathbb S_d(G)$, because $S_v\cup
N_{S_v}=S\cup \tilde N_S$.  As a result $|S_v\cap C|$ must be even
for all $C\in \mathbb C_d^K$ and $S\in \mathbb S_d(G_{v})$. Because
\begin{equation}
(S\cap C_v)\bigtriangleup (S_v\cap
C)=\emptyset\bigtriangleup^{|S\cap N_v|}(v\cap
C)\bigtriangleup^{|v\cap C|}(S\cap N_v)
\end{equation}
is an even set, it follows that $|S\cap C_v|$ must also be even for
all $C_v\in \tilde{\mathbb C}_d^K$. Thus Condition 1 is satisfied.
Thirdly let us suppose that $C_v\bigtriangleup C_v^\prime$ has a
$p$-cover $(\delta,\omega)$ in the new graph $G_{v}$ with $p<d$,
i.e. $C_v\bigtriangleup C_v^\prime=\delta\bigtriangleup \tilde
N_\omega$. Then $(\delta^\prime,\omega^\prime)$ provides a cover for
$C\bigtriangleup C^\prime$ in $G$, i.e. $C\bigtriangleup
C^\prime=\delta^\prime\bigtriangleup N_{\omega^\prime}$ where
$\delta^\prime=\delta\bigtriangleup(\omega\cap N_v)$ and
$\omega^\prime=\omega\bigtriangleup^{m}v$ where
\begin{equation}
m=|\omega\cap N_v|+|v\cap (C\bigtriangleup C^\prime)|
\end{equation}
Since $m+|v\cap \delta|$ is even we obtain
$|\delta^\prime\cup\omega^\prime|=p$. Thus $C\bigtriangleup
C^\prime$ has a $p<d$ cover in graph $G$ which is in contradiction
with Condition 2 of a coding clique. Therefore $C_v\bigtriangleup
C_v^\prime$ cannot be covered by less than $d$ vertices and belongs
to the $d$-uncoverable set of $G_v$, i.e., $C_v\bigtriangleup
C_v^\prime\in \mathbb D_d(G_{v})$ for every $C_v,C^\prime_v\in
\tilde{\mathbb C}^K_d(G)$. Condition 2 is also satisfied. In
addition if $\mathbb C_d^K$ is a coding group of $G$ then
$\tilde{\mathbb C}_d^K$ is a coding group of $G_v$ since
$C_v\bigtriangleup C^\prime_v=(C\bigtriangleup C^\prime)_v$.

Let us denote by $P$ and $\tilde P$ two projectors of the code
subspaces  specified by $C_d^K$ of $G=(V,\Gamma)$ and $\tilde
{\mathbb C}^K_d$ of $G_v=(V,\tilde\Gamma)$ i.e.,
\begin{equation}
P=\sum_{C\in \mathbb
C^K_d}\mZ_C|\Gamma\rangle\langle\Gamma|\mZ_C,\quad \tilde
P=\sum_{C_v\in \tilde{\mathbb
C}^K_d}\mZ_{C_v}|\tilde\Gamma\rangle\langle\tilde\Gamma|\mZ_{C_v}.
\end{equation}
By denoting an LCT as $\mU=\sqrt{-i\mX_v}\prod_{u\in
N_v}\sqrt{i\mZ_{u}}$ we have
$\mU|\Gamma\rangle\langle\Gamma|\mU^\dagger=|\tilde\Gamma\rangle\langle\tilde\Gamma|$
and $\mU \mZ_C\mU^\dagger = \mZ_{C}=\mZ_{C_v}$  if $v\notin C$ and
$\mU \mZ_C\mU^\dagger = i\mZ_{C_v}\mG_v$ if $v\in C$. As a result
$\mU P\mU^\dagger =\tilde P$, i.e., two codes are related to each
other by the LCT $\mU$.

\paragraph{Weight distribution and Classification of all the $[[7,1,3]]$ codes}
Given a $K$ dimensional subspace with projector denoted by $P$ of
$n$ qubit Hilbert space  and an arbitrary set $\omega$ of qubits,
one can build an invariant \cite{lc} under local unitary
transformations as
\begin{equation}
A_\omega=\frac1{K^2}\sum_{{\rm supp}(\mE)=\omega}\left|\mbox{Tr}(\mE
P)\right|^2
\end{equation}
where the summation is taken over all Hermitian Pauli errors that
acting nontrivially on the qubits in $\omega$. If $\omega=\emptyset$
we have $A_0=1$. If we sum over all possible subsets containing $d$
qubits then we obtain the  weight distribution \cite{enum}
$(A_0,A_1,A_2,\ldots,A_n)$  of a code where
\begin{equation}
A_d=\sum_{|\omega|=d}A_\omega, \quad d=0,1,\ldots, n
\end{equation}
are invariant under LUTs and permutations of qubits. Obviously
$\sum_{d=0}^nA_d=2^n/K$ and we neglect $A_0$ and those zero entries
sometimes in the weight distributions. For nonadditive codes $A_d$
may be a fractional. For example the weight distribution of the
nonadditive $(G_{10},24,3)$ code is $((\frac{20}3)_6,35_8)$ meaning
that $A_0=1$, $A_6=20/3$, $A_8=35$, and $A_i=0$ otherwise.

For stabilizer codes $A_d$ is an integer which equals to the number
of the stabilizers with the same weight $d$. The weight of a Pauli
operator is the number of qubits on which it acts nontrivially. For
all the stabilizer codes $[[7,1,3]]$ obtained by searching the
coding groups can have only 10 different weight distributions as
documented in the following table
\begin{center}
{\small\begin{tabular}{@{\extracolsep{0.07cm}}c|c|c|c|c|c|c|c|c|c|c}
     &$W_0$&$W_1$& $W_2$& $W_3$& $W_4$& $W_5$& $W_6$& $W_7$& $W_8$& $W_9$\cr\hline
$A_1$&0& 0& 0& 0& 0& 0& 0& 0& 0& 0\cr $A_2$& 5& 3& 2& 2& 1& 1& 1& 0&
0& 0\cr $A_3$& 0& 0& 0& 0& 2& 0& 0& 2& 0& 0\cr $A_4$& 11& 15& 17& 9
& 7 & 19& 11& 9 & 21& 13\cr $A_5$& 0 & 0 & 0 & 24& 24& 0 & 24& 24& 0
& 24\cr $A_6$& 47& 45& 44& 20& 23& 43& 19& 22& 42& 18\cr $A_7$&0&0
&0 &8 &6 &0 &8 &6 &0 &8\cr
\end{tabular}
}\end{center} By using the weight distributions we can only identify
10 different classes of $[[7,1,3]]$ codes. Further classification is
done by the \textit{frequency analysis} described as follows. Given
$d$ and a subset $S$ of qubits we define
\begin{equation}
F_d(S)=\sum_{\omega\supseteq S,|\omega|=d} A_\omega,\quad
d=1,2,\ldots,n
\end{equation}
to be the frequency of $S$, which is obviously an LU-invariant
quantity. For every $d$ we order all the frequencies $F_d(S)$ of $S$
containing the same number of qubits by their magnitudes. The
resulting ordered series of frequencies is invariant under
permutations of qubits. Any difference between the corresponding
series of two codes will witness their inequivalency. As a result of
frequency analysis we obtain 16 different inequivalent codes and
within each equivalent class all codes can be related to each other
via explicit local Clifford transformations.

\paragraph{Proof of the uniqueness of the code $[[8,3,3]]$}
All the stabilizer codes specified by the coding groups of graphs on
8 vertices have the same weight distribution $(28_6,3_8)$ and 3
stabilizers with full support form a group. Therefore the code must
be pure $(A_1,A_2=0)$ and it is possible to bring two weight 8
generators of the stabilizer to $\mathcal X_V$ and $\mathcal Z_V$ by
LCTs. ($V$ denotes 8 qubits here.) As a result one can always bring
the local operator acting on the first qubit of the remaining three
generators to the identity operator by choosing different
generators. Denote $[G_x|G_z]$ as the check matrix of the remaining
3 generators. The $3\times 8$ matrix $G_{x}$ or $G_{z}$ must have
distinct columns because all single-qubit $\mZ$ or $\mX$ errors can
be corrected and two matrices $G_x$ and $G_z$ cannot have 2 or more
identical columns on the same qubit because all the single-qubit
$\mathcal Y$ errors can be corrected and the code is pure. Thus the
columns of $G_z$ form a map of the the columns of $G_x$ with only
one fix point and this map is unique up to LCTs and permutations
\cite{calder}. As a result the code $[[8,3,3]]$ is unique.

\bibliography{ref4}

\begin{thebibliography}{99}
\bibitem{shor} P.W. Shor, \textit{Phys. Rev. A } \textbf{52}, R2493 (1995).
\bibitem{bdsw} C.H. Bennett, D.P. DiVincenco, J.A. Smolin, and W.K. Wootters, \textit{Phys. Rev. A} {\bf 54},
3824 (1996).
\bibitem{kl} E. Knill and R. Laflamme, \textit{Phys. Rev. A} {\bf 55}, 
900 (1997).
\bibitem{steane} A. Steane, \textit{Phys. Rev. Lett.} {\bf 77}, 793 (1996).
\bibitem{ftc} E. Knill, R. Laflamme, W. H. Zurek, \textit{Science }\textbf{279}, 342 (1998);
D. Gottesman, \textit{Phys. Rev. A} \textbf{57}, 127 (1998).
\bibitem{qkd} C. H. Bennett, G.Brassard, 
\textit{Proceedings of IEEE International
Conference on Computers, Systems, and Signal 
Processing}, 175 (1984); A. K. Ekert, \textit{Phys. Rev. Lett.} \textbf{67}, 661 (1991).
\bibitem{ep} S. Glancy, E. Knill, and H. M. Vasconcelos, 
\textit{Phys. Rev. A} \textbf{74}, 032319 (2006).
\bibitem{gott1} D. Gottesman, \textit{Phys. Rev. A} {\bf 54}, 1862 (1996).%A class of quantum error-correcting codes saturating the quantum hamming bound.
\bibitem{crss} A. R. Calderbank, E. M. Rains, P. W. Shor, and N. J. A. Sloane,
\textit{Phys. Rev. Lett.} {\bf 78}, 405 (1997).%Quantum error correction and orthogonal geometry
\bibitem{calder} A.R. Calderbank, E.M. Rains, P.W. Shor, and
N.J.A. Sloane, \textit{IEEE Trans. Inf. Theory} {\bf 44}, 1369 (1998).%QEC via codes over GF(4)
\bibitem{css} A. R. Calderbank and P.W. Shor, \textit{Phys. Rev. A} \textbf{54}, 1098 (1996); A.M. Steane, \textit{Phys. Rev. A} {\bf 54}, 4741 (1996).
\bibitem{tc1} A.Yu. Kitaev, \textit{Ann. Phys.} (N.Y.) \textbf{303}, 2 (2003).
\bibitem{tc2} H. Bombin and M. A. Martin-Delgado, \textit{Phys. Rev. Lett.} {\bf 97}, 180501 (2006).
\bibitem{eaqecc}T. Brun, I. Devetak, M.-H. Hsieh, \textit{Science} \textbf{314}, 436
(2006).
\bibitem{sdual} L.E. Danielsen and M.G. Parker, \textit{J. Combinatorial
Theory A} {\bf 113}, 1351 (2006).% On the classification of all self-dual additive codes over GF(4) of length up to 12
\bibitem{rains1} E.M. Rains, R. H. Hardin, P.W. Shor, and N.J.A. Sloane,
\textit{Phys. Rev. Lett} {\bf 79}, 953 (1997).
\bibitem{rains2} E.M. Rains, \textit{IEEE Trans. Inf. Theory} \textbf{45}, 266 (1999).
\bibitem{smolin} J.A. Smolin, G. Smith and S. Wehner, arXiv:
quant-ph/0701065.
\bibitem{nonadd} V.P. Roychowdhury and F. Vatan, \textit{Quantum computing and quantum communications 325,
Lecture Notes in Comput. Sci.} 1509, (Springer, Berlin, 1999); arXiv:
quant-ph/9710031.
\bibitem{nonadd2} V. Arvind and K.R. Parthasarathy, arXiv: 
quant-ph/0206174; V. Arvind, P.P. Kurur, and K.R. Parthasarathy,
quant-ph/0210097.
\bibitem{calder2} V. Aggarwal and R. Calderbank, arXiv:cs/0610159.
\bibitem{12in9} S. Yu, Q. Chen, C.H. Lai, and C.H. Oh, arXiv: quant-ph/0704.2122.
\bibitem{zb} A. Cross, G. Smith, J.A. Smolin, and B. Zeng, arXiv: quant-ph/0708.1021.
\bibitem{tanner} R.M. Tanner, \textit{IEEE Trans. Inf. Theory} {\bf IT-27}, 533 (1981).
\bibitem{graph} M. Hein, J. Eisert, and H. J. Briegel,
\textit{Phys. Rev. A} \textbf{69}, 062311 (2004).
\bibitem{werner} D. Schlingemann and R. F. Werner, \textit{Phys. Rev. A} {\bf 65}, 012308 (2002).
\bibitem{schling} D. Schlingemann, \textit{Quant. Inform. Comput.} \textbf{2}, 307
(2002)
\bibitem{grassl} M. Grassl, A. Klappenecker, and M. Rotteler, \textit{IEEE
Int. Symp. Inform. Theory  Proceedings}, \textbf{45} (2002).
\bibitem{oneway} R. Raussendorf and H. J. Briegel, \textit{Phys. Rev. Lett.} \textbf{86},
5188 (2001).
\bibitem{six} P. Walther, et.al., \textit{Nature} \textbf{434}, 169 (2005);
 C.Y. Lu, et.al., \textit{Nature Physics} \textbf{3}, 91 (2007).
\bibitem{LC} M. Van den Nest, J. Dehaene, and B.D. Moor, \textit{Phys. Rev. A} \textbf{69}, 022316 (2004).
\bibitem{cliquer} We have used the clique finding program
\textit{Cliquer} to search for the coding cliques. Ref.: S. Niskanen and P.R.J. \"{O}stergard, \textit{Cliquer User's Guide, Version 1.0}, (Communications Laboratory, Helsinki University of Technology, Espoo, Finland, Tech. Rep. T48, 2003).
\bibitem{enum} P. Shor and R. Laflamme, 
\textit{Phys. Rev. Lett.} {\bf 78}, 1600 (1997);
E.M. Rains, \textit{IEEE Trans. Inf. Theory} {\bf 44}, 1388 (1998).
\bibitem{n-r} R. Cleve, \textit{Phys. Rev. A} {\bf 55}, 4054 (1997).

\bibitem{lc} M. Van den Nest, J. Dehaene, and B.D. Moor,
\textit{Phys. Rev. A} {\bf 70}, 032323 (2004).


\end{thebibliography}

\bibliographystyle{Science}

\end{document}